\documentclass[conference]{IEEEtran}
\usepackage{amsmath,amssymb,amsfonts}
\usepackage{algorithmic}
\usepackage{graphicx}
\usepackage{textcomp}
\usepackage{xcolor}
\usepackage{adjustbox}
\usepackage{gensymb}  
\usepackage{multirow}
\usepackage[utf8]{inputenc}
\usepackage[english]{babel}
\usepackage[top=1.905 cm, bottom=2.54 cm, left=1.5875 cm, right=1.5875 cm]{geometry}

\begin{document}
\title{Impact of Electric Vehicles Botnets on the Power Grid\\}
\author{\IEEEauthorblockN{Omniyah Gul M Khan\IEEEauthorrefmark{1},
Ehab El-Saadany\IEEEauthorrefmark{2}\IEEEauthorrefmark{1},
Amr Youssef\IEEEauthorrefmark{3} and
Mostafa Shaaban\IEEEauthorrefmark{4} }
\IEEEauthorblockA{\IEEEauthorrefmark{1}Electrical and Computer Engineering, University of Waterloo, Canada}
\IEEEauthorblockA{\IEEEauthorrefmark{2}Advanced Power and Energy Center, EECS Department, Khalifa University, UAE \\
and Adjunct Professor, ECE Department, University of Waterloo, Canada}
\IEEEauthorblockA{\IEEEauthorrefmark{3}Concordia Institue for Information Systems Engineering, Concordia University, Canada}
\IEEEauthorblockA{\IEEEauthorrefmark{4}Department of Electrical Engineering, American University of Sharjah, U.A.E}}
\maketitle
\begin{abstract}
The increased penetration of Electric Vehicles (EVs) in the transportation sector has increased the requirement of Fast Charging Direct Current (FCDC) stations to meet customer's speedy charging requirements. However, both charging stations and EVs connection to the communication infrastructure as well as the power grid makes it vulnerable to cyber attacks. In this paper, the vulnerability of the EV charging process is initially studied. We then show how a botnet of compromised EVs and FCDC stations can be utilized to launch cyber attacks on the power grid resulting in an increase in the load at a specific time. The effect of such attacks on the distribution network in terms of line congestion and voltage limit violations is investigated. Moreover, the effect of the botnet of the transmission network is also studied.
Simulation results demonstrate the possibility of line failures, and power outage; and hence, the system's vulnerability to cyber attacks is established.
\end{abstract}

\section{Introduction}
The increased deployment of Electric Vehicles (EVs) has increased the need for public rapid charging stations. Currently, based on the SAE J1772 standard \cite{ref1}, the following three main charging methods are used either residentially, commercially or in public environments: 
\begin{itemize}
    \item Slow Charging-Level 1: It is mainly a residential charger that supplies AC energy to the EV on-board charger at a maximum rate of 16 amps at 120V. Depending on the EV's battery technology, level 1 charging takes 8 to 12 hours to fully charge an empty battery.
    \item Standard Charging-Level 2: It is typically found residentially or commercially supplying AC energy to the EV on-board charger at a maximum rate of 80 amps at 240V. Depending on the EV's battery technology, level 2 charging takes 4 to 6 hours to fully charge an empty battery.
    \item Fast Charging-Level 3: It is typically found commercially or on highways supplying DC energy from an off-board charger to the EV. Most common forms of level 3 chargers currently provide 50kW charge per hour. Tesla level 3 chargers, on the other hand, provides 120kW of charge per hour.
\end{itemize}
As EVs are becoming more popular day by day, they are becoming more affordable to middle and lower class households. However, not all EVs consumers own their own residential property to charge their cars. Instead, many of them may live in more dense environments that would require the installation of public level 2 or level 3 charging stations to satisfy their charging needs. Moreover, to attract more people into driving EVs, the future would see the deployment of a large scale fast charging Direct Current (FCDC) stations that provides speedy charging. However, both EVs and charging stations are considered as \textit{Internet of Things} (IoT) devices due to their interconnection via the Internet. Hence, cyber security of EVs and charging stations play an important role in their design and integration in the electric grid. Moreover, relevant EV charging communication standards are in its infancy and not open enough forcing security by obscurity.

Cyber attackers can create a network of infected EVs and charging stations, referred to as a botnet, to cause a larger effect on the power grid (A botnet is a network of controllable devices that are remotely controlled by the botmaster to commit cyber attacks \cite{botnet}). Charging of EVs already is of concern in terms of unconsidered increase in the load especially during peak time. An EV botnet can cause a large effect on the grid if the attacker decides to create a scenario utilizing the botnet to further increase drastically the load on the grid, most probably leading to outages. The effect of such an EV botnet attack on the distribution and transmission networks will be studied in this paper.  

Studies on cyber security of modern cars essentially started around 2010. In \cite{ref7}, the external attack surface of cars was studied. Different attacks were implemented compromising the Electronic Control Unit (ECU) of a car. Attackers were able to utilize the vulnerable car's external I/O interfaces to remotely control various vehicular functions. In \cite{ref8}, a wireless attack was implemented using a connected vehicle where the driver's phone was connected to the vehicle's Controller Area Network (CAN). Initially, the attacker was assumed to have acquired access to the Car Area Network (CAN) data frame to control the vehicle. Moreover, the attacker was assumed to have published a malicious app via which the attacker could control the driver's phone when the driver downloads it. Moreover, based on \cite{ref14}, charging station infrastructures are open for potential cyber attacks as they are unmanned and located in remote areas occasionally. On the other hand, \cite{ref9} and \cite{ref11} demonstrated that an IoT botnet of high wattage devices provides attackers the ability to manipulate the demand on the grid resulting in power outages and/or blackouts. Soltan et al. \cite{ref10} provide methods to prevent line failures due to such demand manipulation. However, to the best of our knowledge, no previous work has been done in studying the effect of a botnet made up of EVs and charging stations on the transmission and distribution networks.

The main contributions of this paper are two-fold: $(1)$ assess the vulnerability of the EV charging process and $(2)$ simulate a cyber attack on the distribution network and transmission network and study its corresponding effect on congestion, line failures, and outage. 
\section{Vulnerability Analysis of EV charging}
The basic characteristics that should be present in any cyber security system are represented by the CIA triad \cite{ref12} which symbolizes confidentiality, integrity, and availability. Confidentiality is needed to ensure restricted and authorized access to sensitive information. Integrity, on the other hand, refers to the assurance that information is authentic and not corrupted. Finally, availability is the guarantee that authorized users have access to required information and services at all times. The success of EV charging, hence, depends on the CIA of the cyber-network which can pose significant threats to the smart grid if security vulnerabilities are not addressed.

\begin{figure} [h]
\centering
  \includegraphics[trim={0cm 0cm 0cm 0cm}, scale=0.45]{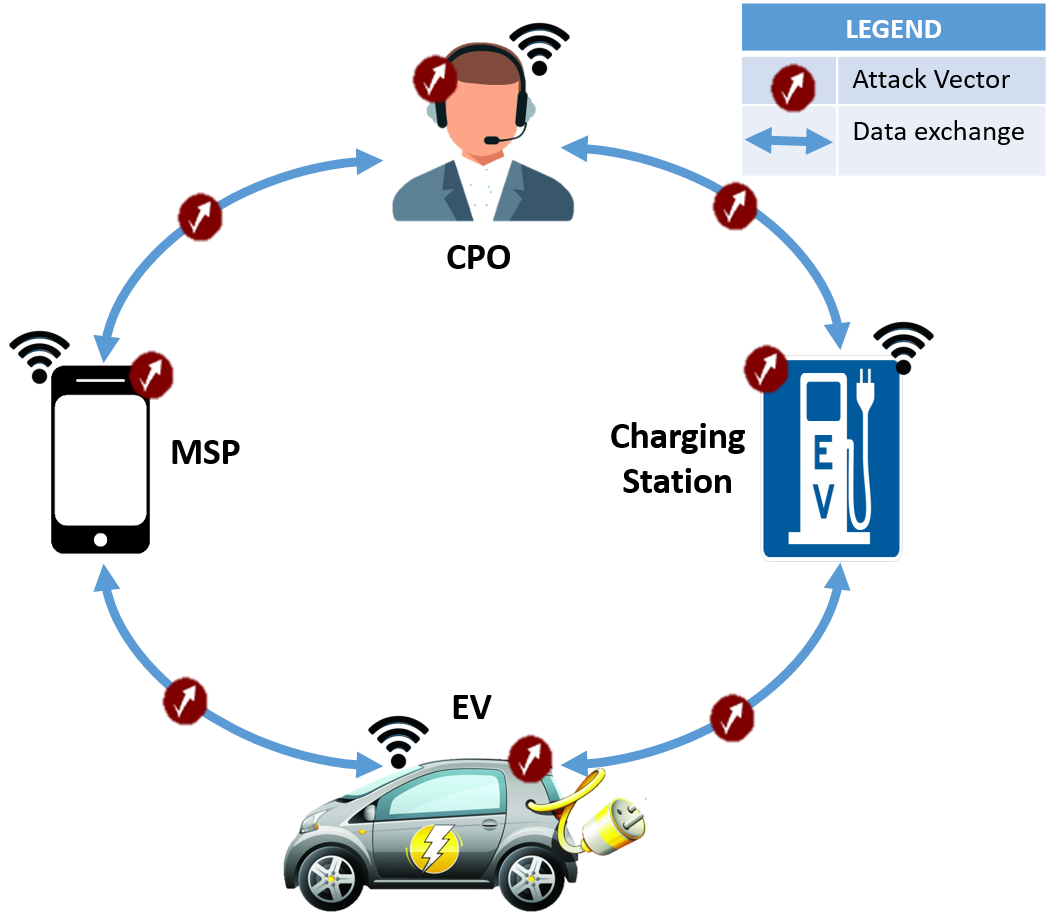}
  \caption{Cyber vulnerabilities in the EV charging process}
  \label{fig:fig1}
\end{figure}

Utilizing an EV's physical and communication connection to a charging station, previous studies have revealed several sources of vulnerabilities that a cyber attacker can utilize in the process of charging an EV \cite{ref13,ref14,ref15}. These vulnerabilities have been summarized as follows (also see Figure \ref{fig:fig1}):
\begin{itemize}
    \item When needing to refuel, an EV user communicates with the Mobility Service Provider (MSP), that he/she has a contract with, through a mobile application to determine the closest charging station that is available. Attacking the MSP can allow access to personal EV driver and vehicle information, direct EV drivers to busy charging stations, show the unavailability of desired charging stations forcing EV owners to drive to more expensive charging options (FCDC), or affect the billing and payment for consumed energy.
    \item Information is exchanged between the EV and the charging station through the connector ports of the charging cable. The State of Charge (SOC) is initially communicated by the EV to the charging station along with the EV battery’s voltage ranges and ampacity \cite{ref5}.  If the EV is infected with a malware, the charging station could also get infected and the malware can spread to other EVs that refuel using the same charging station in the future. Mobility of EVs can also be used to spread the malware to other charging stations.
    \item Attacking the charging station itself can cause the EV not to get charged to the desired SOC, i.e. the attacker can stop charging the EV. The attacker can also cause the EV to get charged free of charge. Also, the attacker can  ignore the ampacity limit of the EV battery and charge at a rate beyond the capacity of the EV causing intentional damage to the EV. 
    \item The Charge Point Operator (CPO) is responsible for operating and maintaining a collection of charging stations. When a user arrives at the charging station, the MSP verifies the user and the CPO is then capable of starting the charging process. Attacking the CPO can result in providing the wrong amount of charge to the EV or even stopping the charging process without achieving the user's requirement,
    \item A cyber attacker can implement a Man in the Middle attack and intercept or even change information exchanged between the EV and the charging station, EV and the MSP, MSP and the CPO, or the CPO and the charging station \cite{ref16}.
\end{itemize}
Utilizing any of these vulnerabilities can have devastating effects. In this paper, charging stations and EVs are compromised to form a botnet that increases the grid load. However, attacks such as Denial of service attacks, electricity theft and privacy infringement are not considered in this paper \cite{rerrr} \cite{rerrr2}.

\section{Effect of  EV Botnets on Distribution Network}
\begin{figure}[htbp]
\centering
  \includegraphics[trim={0cm 0cm 0cm 0cm}, width=0.45\textwidth]{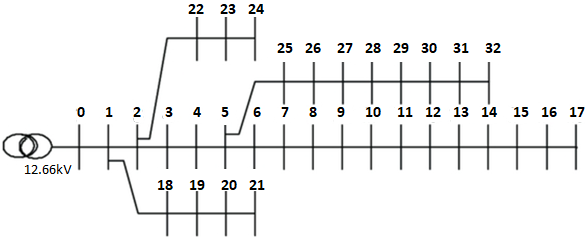}
  \caption{IEEE 33 bus system used as a distribution network case study \cite{ref17}}
  \label{fig:fig2}
\end{figure}

To study the effect of an EV botnet on the distribution network, the IEEE 33 bus network \cite{ref17}, illustrated in Figure \ref{fig:fig2}, is utilized. The non-flexible base load profiles of the 32 load buses are designed to be equivalent to the default loading of the IEEE 33 bus system \cite{ref17}. All the load buses, except bus 24, are assumed to be residential, consisting of EV and Heat Pumps (HP) \cite{b22} as flexible loads. Having the largest base load, bus 24 represents a commercial bus consisting of an EV fast charging parking lot that could accommodate fifty EVs. Data obtained from Toronto parking authority \cite{TPA} representing arrival and departure times of EVs in a commercial area parking lot was used to simulate the availability of EVs in the parking lot on bus 24. The initial state of charge (SOC) of each EV is randomized to be uniformly distributed between $20\%$ and $30\%$ and it is assumed that the owners would want the car to be completely charged on departure. The various parameters needed for simulating the operation of the HP and the EV, and performing power flow analysis are listed in Table \ref{table:table2}.
\setlength{\textfloatsep}{5pt}
\begin{table}[t!]
\begin{center}
\caption{Case Study parameters}
\begin{tabular}{ |c|c| } 
 \hline
 \textbf{Variable} & \textbf{Value} \\
 \hline  \hline
 Coefficient of Performance (COP) & 2.2 \\
 Residential House Areas & 1500-2500 sq.ft\\
 House Temperature Range & 18-24\degree C\\
 EV battery size & 36 kWh\\
 Maximum Residential EV charging power & 11 kW\\
Maximum Fast Charging EV power & 50 kW\\
 Initial SOC of EV & 0.2-0.3\\
 Power flow limit of branch 1-2 & 7.91 MW\\
 Upper voltage limit   & 1.05 pu\\
 Lower voltage limit & 0.95 pu\\
 \hline
\end{tabular}
\label{table:table2}
\end{center}
\end{table}

\begin{figure}[!h]
\centering
  \includegraphics[trim={0cm 0cm 0cm 0cm}, width=0.48\textwidth]{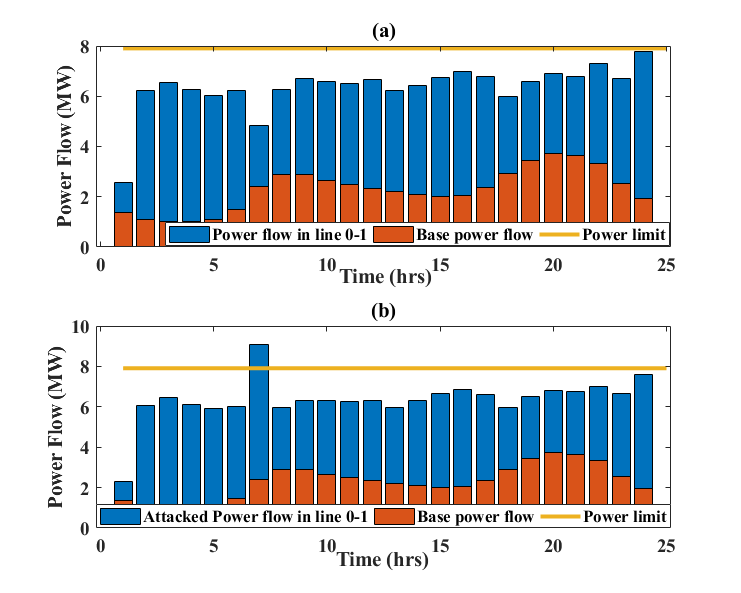}
  \caption{Impact of an EV botnet attack on the loading of line 0-1 where (a) represents the normal operation power flow and (b) represents the power flow as a result of the cyber attack}
  \label{fig:fig3}
\end{figure}

Moreover, fast charging stations are assumed to be placed on some of the residential buses for consumers living in multi-dwelling units lacking the space required for home charging stations. MATPOWER \cite{mat} was used to simulate the base case of the IEEE 33 bus network to determine the weakest point in the network in terms of the largest voltage drop. Bus 17 followed by bus 16 were observed to have the lowest bus voltages after performing load flow analysis. Hence, assuming that the attacker has knowledge of the network she is attacking, fifty FCDC stations are assumed to be located at these two buses. Figure \ref{fig:fig3}(a) illustrates the amount of power flow with and without the flexible load in branch 0-1 of the IEEE 33 system. As observed, there is no congestion as not all the FCDC are operational at the same instant of time. However, when the attacker utilizes the EV botnet and directs all EVs requiring to be charged to the FCDC charging stations at time 07:00, the demand for electricity increases as all the FCDC stations are operational. Such an attack is a type of load altering attack \cite{laa} which results in congestion in the grid as observed in Figure \ref{fig:fig3}(b) where the power flowing in branch 0-1 is exceeding its capacity. Distribution Network Operator (DNO) would require to immediately load shed or line failure of branch 0-1 occurs causing a power outage.

\begin{figure}[!h]
\centering
  \includegraphics[trim={0cm 0cm 0cm 0cm}, width=0.48\textwidth]{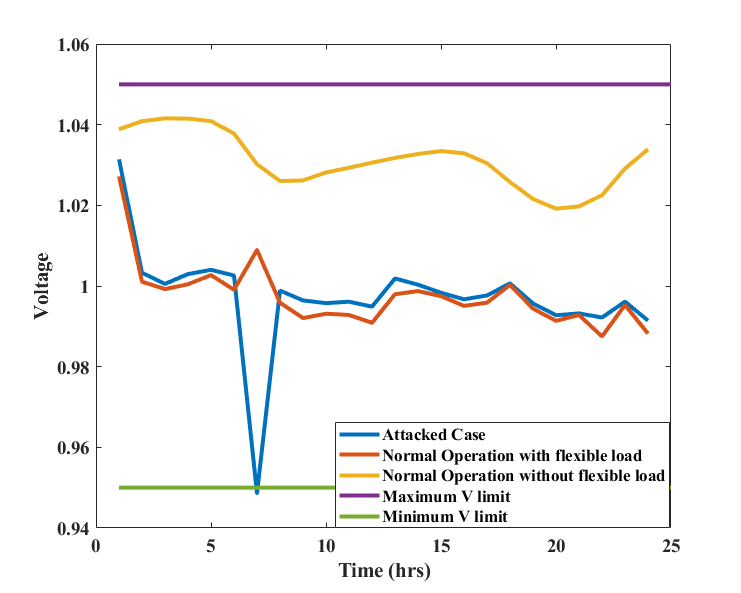}
  \caption{Impact of an EV botnet cyber attack on the minimum bus voltages}
  \label{fig:fig4}
\end{figure}

Since the FCDC points were located at the two weakest voltage points at bus 16 and 17, the effect on voltage can be clearly seen in Figure \ref{fig:fig4}. Due to the cyber attack at time 07:00, the minimum voltages at that time decreases to below the NEC 5\% voltage drop recommendation. Hence, either voltage support methods need to be implemented in the network which would incur cost or the voltage that is supplied to the consumer would not be at the rated voltage causing improper, erratic or no operation of the loads and eventually damage to the equipment. Moreover, an excess voltage drop is a concern due to the potential fire hazard at high resistance connections. 

\section{Effect of  EV Botnets on Transmission Network}
\begin{figure}[!h]
\centering
  \includegraphics[trim={0cm 0cm 0cm 0cm}, width=0.35\textwidth]{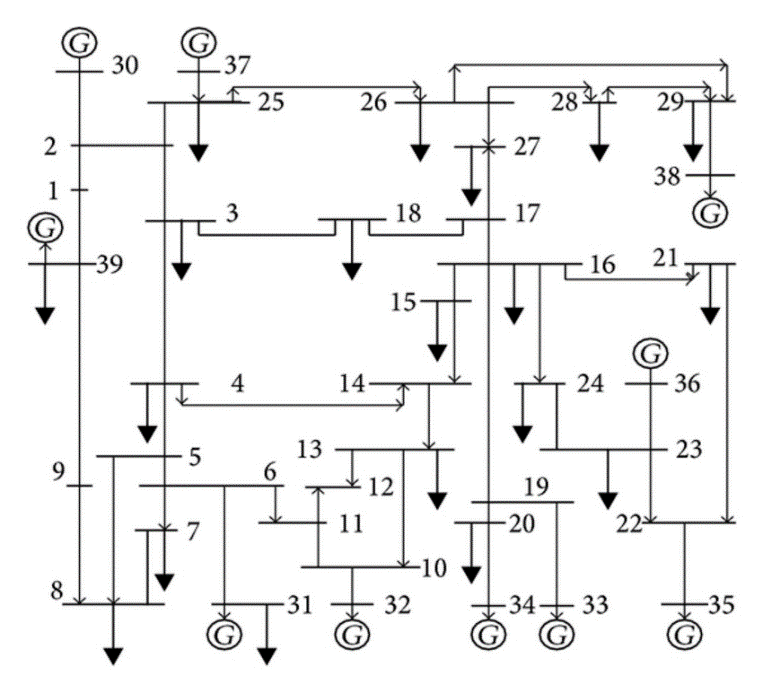}
  \caption{IEEE 39 bus system used as a transmission network case study \cite{ref39bus}}
  \label{fig:fig5}
\end{figure}

A load altering attack is generally observed at the distribution network level. However, the impact of such an attack can be comprehended by studying the system at the transmission level. Hence, to study the effect of an EV botnet on the transmission network, the IEEE 39 \cite{ref39bus} bus system was used which consists of 39 buses, 10 generators and 46 branches as illustrated in Figure \ref{fig:fig5}.

\begin{figure}[!b]
\centering
  \includegraphics[trim={0cm 0cm 0cm 0cm}, width=0.48\textwidth]{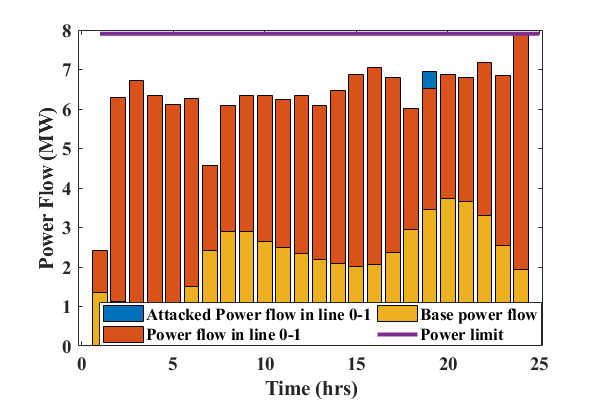}
  \caption{Impact of an EV botnet consisting of 5 FCDC stations on bus 16, 17 and 24 on the loading of line 0-1 }
  \label{fig:fig6}
\end{figure}

The IEEE 39 bus system has 19 load buses. To simulate a cyber attack that utilizes an EV botnet, the load on these buses is increased by 5\%. For example, bus 20 has a 680MW base load. Hence, an increase of 5\% corresponds to an increase of 34MW which represents 680 FCDC points being operational at the same time to charge EVs. In order  not to trigger the protection system in the distribution network, this increase in the load buses should not cause any congestion in their respective networks. Bus 9 of the IEEE 39 bus system has a load of 6.5MW which corresponds to the load at time 19:00 of the IEEE 33 bus system explained in the previous section. Hence, the IEEE 33 bus system with loads at time 19:00 were used to simulate the load on bus 9 of the IEEE 39 bus system. Initially, in the absence of an attack, 5 FCDC points on each of the buses 24, 17 and 16 were simulated in which EVs were being charged based on the data obtained from the Toronto Parking Authority \cite{TPA}.  Load flow analysis was performed to obtain the power flow in branch 0-1 and observe if there will be any congestion. A cyber attack is then triggered to cause all the 5 FCDC points to be charging at time 19:00. Figure \ref{fig:fig6} depicts the absence of congestion even when the attack happens at 19:00 causing an increase in the load by 6.5\%. Hence, no protection system at the distribution level is triggered. 

When the load at 19:00 is then used to study the effect of such attacks on the transmission system, on performing load flow analysis, branch 24 linking bus 15 and 14 is observed to be overloaded. This branch is deactivated, and the network is still complete and no isolated islands are observed to exist. Load flow analysis is re-performed resulting in overloading branch 7 linking bus 18 and 3. This branch is also deactivated, and the network is still complete and no isolated islands are observed to exist. Load flow analysis is repeated resulting in overloading branches 31 linking bus 27 and 17, and 40 linking buses 26 and 25. This causes the creation of three islands: island 1 which consists of 23 branches, 5 generators and 20 buses. Island 2 consists of 14 buses, 4 generators and 14 buses. Island 3 consists of 5 branches, 1 generator and 5 buses. Moreover, even though we have lines being deactivated due to being overloaded, the system is still capable of providing all the loads. Hence, no outage is observed.

\begin{figure}[!h]
\centering
  \includegraphics[trim={0cm 0cm 0cm 0cm}, width=0.48\textwidth]{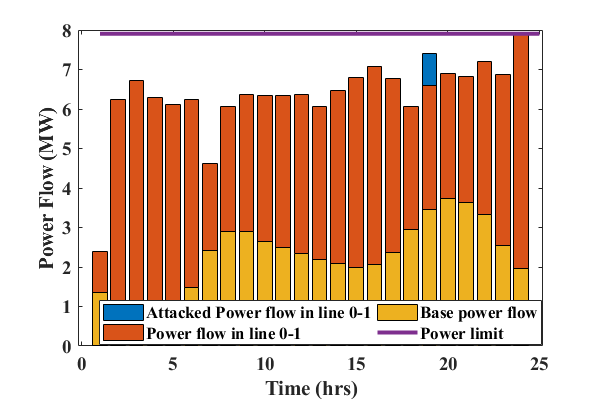}
  \caption{Impact of an EV botnet consisting of 10 FCDC stations on bus 16, 17 and 24 on the loading of line 0-1}
  \label{fig:fig7}
\end{figure}

However, increasing the load on all the load buses by 10\% causes an outage. Initially, it was ensured that this increase in the load buses does not cause any congestion in the distribution system triggering its protection system. The same process explained earlier was repeated utilizing the IEEE 33 bus system as a load for bus 9 of the IEEE 39 bus system having a load of 6.5MW. In the absence of an attack, 10 FCDC points on each of the buses 24, 17 and 16 were simulated in which EVs were being charged based on the data obtained from the Toronto Parking Authority. Load flow analysis was performed to determine the power flowing in branch 0-1 and observe if there will be any congestion. A cyber attack is then triggered to cause all the 10 FCDC points to be charging at time 19:00. Figure \ref{fig:fig7} illustrates the absence of congestion in the distribution network even when the attack happens at 19:00 causing an increase in the load by 11.8\%. Hence, no protection system at the distribution level is triggered. 

\begin{figure}[!h]
\centering
  \includegraphics[trim={0cm 0cm 0cm 0cm}, width=0.35\textwidth]{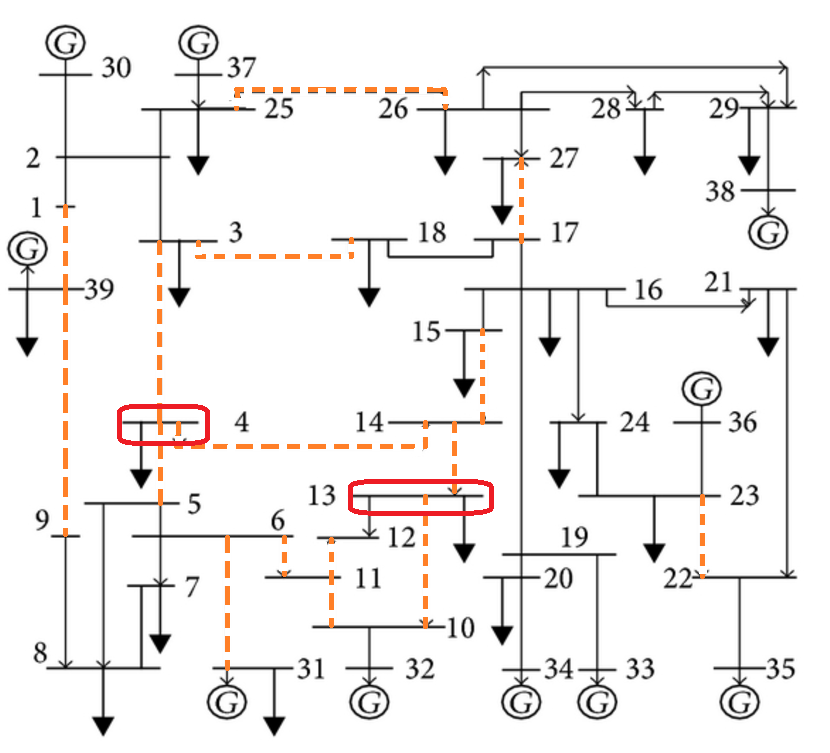}
  \caption{Two dead load buses formed as a result of overload in lines due to the botnet cyber attack increasing the load in all buses by 10\%}
  \label{fig:fig8}
\end{figure}

When the load at 19:00 is then used to study the effect of such attacks on the transmission system, 4 branches are observed to be overloaded. These branches are deactivated, and the network is still complete and no isolated islands are observed to exist. Load flow analysis is re-performed resulting in the overload of 9 branches which when deactivated causes the creation of nine islands. This process continues resulting in 16 branches being overloaded and an outage of 559.4MW (represents 8\% of the original load) since load bus 13 and 4 are isolated and has no way to be provided. Figure \ref{fig:fig8} illustrates the deactivated branches using dotted lines and the red boxes represents the isolated load buses.

\section{Conclusions}
The effect of a cyber attacker exploiting a botnet comprising of EVs and FCDC stations to attack the grid was investigated in this paper. The vulnerability of the EV charging process was studied initially and the IEEE 33 bus system was used to demonstrate the effect of increasing the load using the EV botnet on the distribution network causing line congestion and voltage limit violations. Moreover, the attacker could increase the load such that the distribution network's protection system is not triggered yet the effect of the attack is observed in the transmission network in terms of outages. The security of EVs and FCDC stations is hence, vital for the safe and secure operation of the power grid. Further studies investigating the effect of a coordinated attack on multiple charging stations on the system's voltage and frequency needs to be performed. Moreover, preventive methods that detect and protects the system against such attacks need to be developed. 
\nocite{ref14}
\bibliography{main}

\end{document}